\documentclass[%
reprint,
 amsmath,amssymb,
 aps,
]{revtex4-2}
\usepackage{graphicx}
\usepackage{dcolumn}
\usepackage{bm}

\usepackage{color}
\usepackage{ulem}
\usepackage{orcidlink}
\usepackage{hyperref}
\hypersetup
{
hypertex=true,
colorlinks=true,
linkcolor=blue,
filecolor=blue,
urlcolor=blue,
citecolor=blue,
breaklinks=true
}
\begin{document}
\preprint{APS/123-QED}
\title{Significantly enhanced detectability of dark photons with a steady-state excited microwave cavity}
%
\author{S. R. He\orcidlink{https://orcid.org/0000-0002-2837-3410}}

\author{L. Gao \orcidlink{https://orcid.org/0000-0002-2837-3410}}

\author{P. H. Ouyang\orcidlink{https://orcid.org/0009-0000-5379-8552}}

\author{H. Zheng \orcidlink{https://orcid.org/0000-0002-2837-3410}}

\author{X. N. Feng
\orcidlink{https://orcid.org/0000-0002-2837-3410}}

\author{L. F. Wei
\orcidlink{https://orcid.org/0000-0003-1533-1550}}
\email{lfwei@swjtu.edu.cn}
\affiliation{Information Quantum Technology Laboratory, School of Information Science and Technology, Southwest Jiaotong University, Chengdu 610031, China}

\begin{abstract}
The resonant cavity system has been widely used to search for the electromagnetic response of dark photons, although its achievable detection sensitivity remains at a relatively low level. In this letter, we propose a feasible approach to significantly improve its achievable detection sensitivity by enhancing the detectability of the dark photon-photon dynamical effect, assisted with the steady-state excitation of the target mode in the cavity. 
Unlike in almost all the previous detection schemes, wherein where the cavity modes are kept in vacuum (and thus only the second-order energy signals can be detected), here the pre-excited steady-state field in the cavity can be used to achieve the coherent amplification of the dark photon response signal, thereby obtaining detectable first-order (rather than the conventional second-order) energy response signals of dark photons. Although the phase of the dark photon field and thus its electromagnetic response signal is stochastic, the amplitude of such a first-order energy response power signal can still be extracted by using mature IQ demodulation technology. As a consequence, we argue that, even considering the influence of the shot noise of the pre-excited steady-state field, the achievable detection sensitivity of this in-situ enhancement detectability, based on the steady-state excitation signal of the target mode, is still at least one order of magnitude higher than those of the current resonant cavity experiments with the same Q-quality factors. Based on existing microwave cavity and weak signal demodulation detection technologies, the feasibility of such a significantly enhanced detectability scheme is also discussed.
\end{abstract}
\maketitle

{\it Introduction---}
Until now, a series of collider experiments ~\cite{2018RA} and fixed-target scattering experiments~\cite{2025TL,2017IMB}have been utilized to search for the particle-type dark photons with the masses being greater than twice the electron mass. Also, various quantum sensing detection techniques ~\cite{2025YF,2022DA,2025HF,2019VVF,2020AB,2021AVD,2023SC,2023JG,2024AA,2025SRH,2025SC,2024AI,2024SC,2025HS,2025SC} have been developed for the ultra-light wave-type dark photon dark matter whose mass is at the submeV level. In particular, the microwave cavity resonance detection of dark photon-photon kinetic mixing effects for the masses in meV or lower range has always been highly anticipated~\cite{2012PA,2021AC,2025YFC,2022RC,2024APEX,2024ZXT,2024RK,2025YY,2024RC,2025YJZ}.

Although all the experiments conducted so far have yielded the “zero results” detection, the constraints on the value of the dark photon-photon mixing parameter have been continuously lowered~\cite{2021AC,2024ZXT}. The core principle of this widely discussed detection method is that the dark photons entering a resonant cavity can dynamically mix with the photons of the vacuum of the target mode in the cavity and then transfer its energy into the target mode field, thereby inducing a detectable photon signal of that mode in the cavity ~\cite{2021AC,2022RC,2024APEX,2024RK,2024ZXT,2025YY,2024RC,2025YJZ}. In principle, the detection sensitivity achieved could be further improved by using a superconducting radio-frequency cavity with the higher quality factor~\cite{2024ZXT,2024RC}; however, it is essentially limited because its detection is just the extremely weak second-order energy response signal. Therefore, finding a new detection scheme for the dark photon-photon dynamic mixing effect detection, which could replace the above second-order energy converted effect detection, has become a very urgent task.

In this letter, we propose an approach, shown schematically in Fig.~1, to update the conventional microwave cavity detection scheme for significantly improve the detectability of the electromagnetic response of dark photons, even if the quality factor of the cavity is unchanged. In the present configuration, the target electromagnetic mode inside the cavity is no longer prepared at its vacuum state, but in its steady state excitation (which is served as a seed field) under the drive of the outside externally coherent source. Therefore, what dynamically mixes with the dark photon entering the cavity is no longer only the vacuum state of the target mode but also its excited seed field, which might generate a first-order energy response signal of dark photon energy conversion. Unlike only the signal whose power is proportional to the square of the mixing parameter~\cite{2012PA,2021AC} could be detected previously, here the dark photon (indicated by the blue arrow in the figure) can dynamically mix with the cavity photons in the steady-state excited mode (shown as the red wavy line in the figure) and thus a significantly enhanced first-order energy response signal, whose power is proportional to the first power of the mixing parameter, can be generated for detection. 
\begin{figure}[htbp]
\centering
\includegraphics[width=7cm]{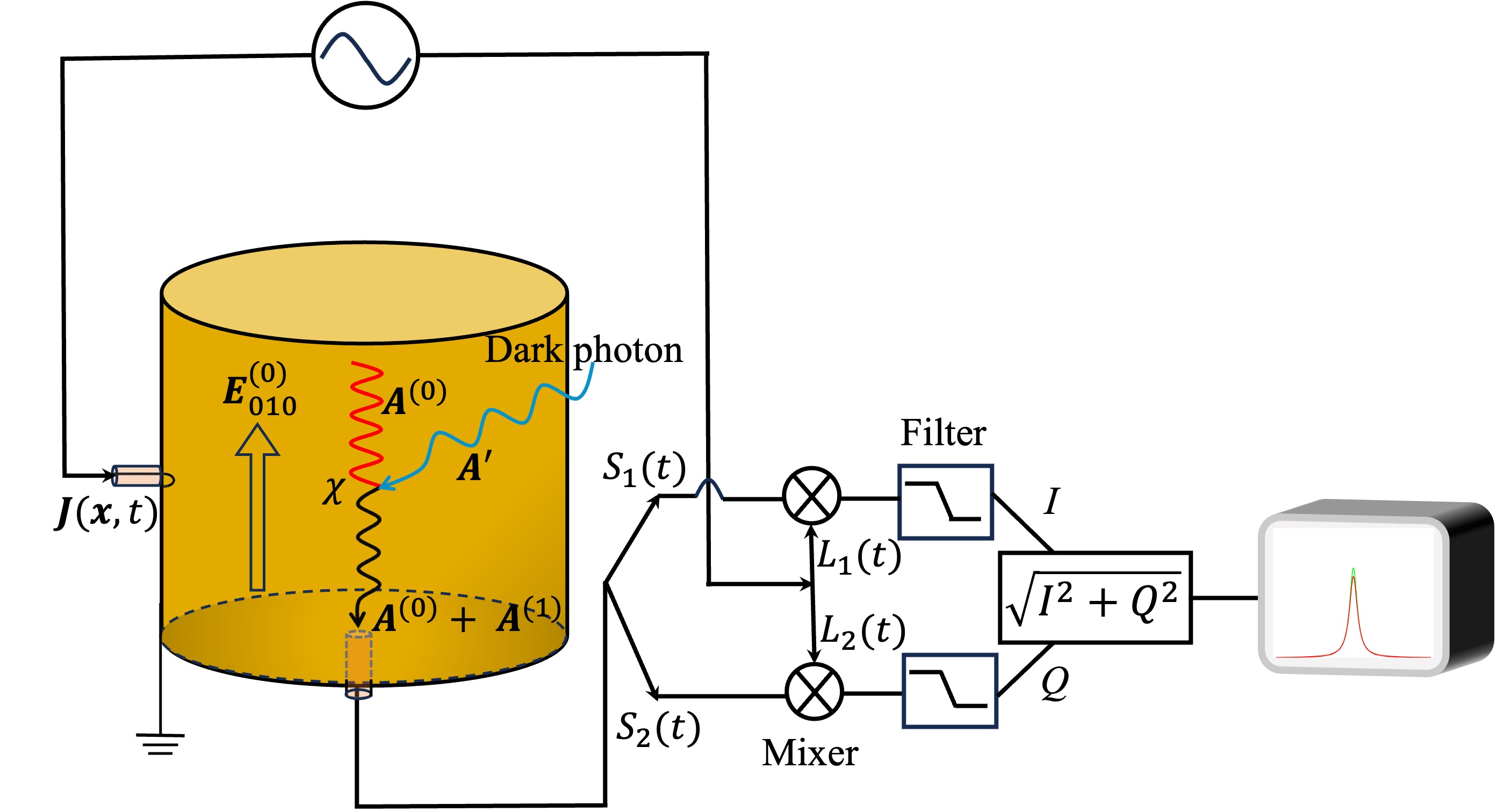}
\caption{A schematic illustration of the upgraded detector for probing the dynamical mixing effect between the dark photon $\boldsymbol{A}'$ and the steady-state seed field $\boldsymbol{A}^{(0)}$ excited by the outside source $\boldsymbol{J}(\boldsymbol{x},t)$. The signal of $\boldsymbol{A}^{(0)}+\boldsymbol{A}^{(1)}$ is subsequently sent to the IQ modulation for detection.}
\end{figure}
Although the time average of such a first-order energy response signal is zero (as the phase information of the dark photon is random) and thus cannot be direct power detection, the amplitude of such a signal can still be detected by using well-known IQ demodulation techniques~\cite{2024QY,2024YL}. Therefore, the upgraded scheme proposed here can significantly increase the detectability of dark photons for the cavity with the same quality factor, and the relevant detection sensitivity can be improved by at least an order of magnitude or more. 

{\it Model and Method---}
Generally, the dynamical mixing between dark photons and photons can be described by the following Lagrangian~\cite{2019VVF,2023JG}:
\begin{equation}
\begin{aligned}\mathcal{L}\supset&-\frac{1}{4\mu_0}F_{\mu\nu}F^{\mu\nu}+J^{\mu}A_{\mu}-\frac{1}{4\mu_0}F'_{\mu\nu}F'^{\mu\nu}\\&+\frac{m_{d}^2\varepsilon_0}{2}A_\mu^{\prime}A^{\prime\mu}-m_{d}^2\varepsilon_0\chi A_{\mu}A^{\prime\mu},
\end{aligned}
\end{equation}
where $A_\nu$ and $A'_\nu$ denote the gauge four-vectors of the ordinary photons and the dark photons, respectively; $F_{\mu\nu}=\partial_\mu A_\nu-\partial_\nu A_\mu$ and $F'_{\mu\nu}=\partial_\mu A'_\nu-\partial_\nu A'_\mu$ are the corresponding electromagnetic field strength tensors; $J^{\mu}$ is the four current density vector of the ordinary electromagnetic field; $\varepsilon_0$ is the vacuum permittivity; $\chi$ is the kinetic mixing parameter (i.e., the detection sensitivity parameter) and $m_d$ the dark photon mass, expressed in angular frequency units. From Eq.~(1), the equation of motion for the ordinary electromagnetic field in the presence of dark photon mixing can be written as
\begin{equation}
\left(\frac{1}{c^{2}}\frac{\partial^{2}}{\partial t^{2}}-\nabla^{2}\right)A^{\nu}=\mu_{0}J^{\nu}+\boldsymbol{J}_{\mathrm{eff}},
\end{equation}
where $c$ is the speed of light, $\mu_0$ is the vacuum permeability, and $\boldsymbol{J}_{\mathrm{eff}}=m_{d}^2\chi \boldsymbol{A'}(\boldsymbol{x},t)/c^2$ is the effective current induced by dynamical mixing with the dark photon with $\boldsymbol{A'}(\boldsymbol{x},t)$ being the vector potential associated with the dark photon field~\cite{2025YFC}. By solving Eq.~(2), one obtains the electric field strength of the ordinary electromagnetic field inside the cavity, $\boldsymbol{E}(\boldsymbol{x},t)=-\partial{\boldsymbol{A}(\boldsymbol{x},t)/\partial t}$, and hence the detectable power of the electromagnetic signal in the cavity reads $P\propto |\boldsymbol{E}(\boldsymbol{x},t)|^2$.

Formally, the vector potential of the electromagnetic field inside the cavity can be generally expressed as $\boldsymbol{A}(\boldsymbol{x},t)=\boldsymbol{A}^{(0)}(\boldsymbol{x},t)+\boldsymbol{A}^{(1)}(\boldsymbol{x},t)+\mathcal{O}^{(2)}(\boldsymbol{x},t)$. Here, $\boldsymbol{x}$ and $t$ denote the spatial position vector and time, respectively. Noted that $\boldsymbol{A}^{(0)}$ represents the ordinary electromagnetic field inside the resonant cavity (i.e., the zeroth-order response of dark photons), which satisfies the usual driven dissipative equation:
\begin{equation}
\left(\frac{1}{c^{2}}\frac{\partial^{2}}{\partial t^{2}}-\nabla^{2}+\frac{\omega_j}{Qc^2}\frac{d}{dt}\right)\boldsymbol{A}^{(0)}(\boldsymbol{x},t)=\mu_0\boldsymbol{J}.
\end{equation}
Here, $\boldsymbol{J}$ denotes the ordinary electromagnetic current, while $\omega_j$ and $Q$ are the eigenfrequency and the corresponding quality factor of the $j$th mode of the resonant cavity, respectively.
By contrast, the first-order response $\boldsymbol{A}^{(1)}$ of dark photons is described by
\begin{equation}
\left(\frac{1}{c^{2}}\frac{\partial^{2}}{\partial t^{2}}-\nabla^{2}+\frac{\omega_j}{Qc^2}\frac{d}{dt}\right)\boldsymbol{A}^{(1)}(\boldsymbol{x},t)=\boldsymbol{J}_{eff},
\end{equation}
whereas the higher-order correction term $\mathcal{O}^{(2)}$ is exceedingly weak and thus can be safely neglected.

Extensive previous studies (see, e.g., Refs.~\cite{2012PA,2021AC,2025YFC,2022RC,2024APEX,2024ZXT,2024RK,2025YY,2024RC,2025YJZ}) have focused on detecting the response of the intra-cavity vacuum electromagnetic field to dark photons by setting $\boldsymbol{J}=0$ in Eq.~(3). In this case, the zero- and first-order steady state resonant excitation of the typical, i.e, $\mathrm{TM}_{010}$-mode  (with $\omega_0=m_d$), induced by the effective dark photon current, reads $\langle{\boldsymbol {E}^{(0)}_{010}}\rangle=0$~\cite{2012PA}, and
\begin{equation}
\tilde{E}_{\hat{z},\mathrm{ss}}^{(1)}(r,t)
=\chi\mathbb{B}e^{-i(\omega_0t+\phi_d)},
\end{equation}
with $\langle\cdot\rangle$ being the statistical average and $\mathbb{B}=2 Qc\sqrt{2\rho_d\mu_0}\mathbb{J}_0\left(\frac{x_{01}}{a}r\right)\cos\theta/[x_{01}\mathbb{J}_1(x_{01})]$. Here, $\rho_d$ and $\phi_d$ denote the dark photon density and stochastic phase, respectively. $\mathbb{J}_0(\cdot)$ and $\mathbb{J}_1(\cdot)$ are the zeroth- and first-order Bessel functions of the first kind, respectively. $x_{01}\approx 2.4048$ is the first root of the Bessel function $\mathbb{J}_0(\cdot)$. $r$ is the radial coordinate, $a$ is the radius of the cylindrical resonant cavity. $\theta=\arccos{\langle\hat{z},\hat{\boldsymbol{n}}\rangle}$ denotes the angle between the dark photon polarization direction $\hat{\boldsymbol{n}}$ and the $\hat{z}$ direction (i.e., the cavity axis) of the cylindrical resonant cavity. Although the phase of this first-order response electric field naturally carries the stochastic phase of the dark photon field, it can still produce a detectable power signal (see Appendix A for details):
\begin{equation}
\begin{aligned}
\langle P_{\mathrm{out}}^{(2)}\rangle\approx&\kappa\frac{\omega_0}{Q}\int_V\left(\frac{1}{2}\varepsilon_{0}\Big|\tilde{E}_{\hat{z},\mathrm{ss}}^{(1)}(r,t)\Big|^{2}\right)dV\\
=&\chi^2\frac{4\kappa\omega_0Q\rho_d\pi h a^2\cos^2\theta}{x^2_{01}},
\end{aligned}
\end{equation}
where $\kappa$ is the attenuation coefficient at the coupling port of the resonant cavity, and $h$ is the height of the cylindrical resonant cavity. Notably, $\langle P_{\mathrm{out}}^{(2)}\rangle$ is independent of the stochastic phase of the dark photon field. However, it is a second-order energy response signal, since the detectable power scales as the square of the kinetic mixing parameter $\chi$, making it intrinsically too weak for efficient detection. In other words, conventional cavity-based dark photon detection schemes that rely on this second-order energy response signal suffer from inherently low detection sensitivity and therefore require substantial improvement.

Below, we propose an upgraded version of the conventional microwave cavity resonant detection method to significantly enhance the detectability of dark photons. The basic idea is as follows:

i) Pre-excitation of a cavity mode. Without loss of generality, let the $\mathrm{TM}_{010}$ mode can be excited by using $\boldsymbol{J}\neq 0$. Consequently, the steady-state solution to Eq.~(3) reads:
\begin{equation}
\tilde{E}^{(0)}_{\hat{z},\mathrm{ss}}(r,t)=\mathbb{A}e^{-i\omega_0 t},\mathbb{A}=E_0 \mathbb{J}_0(x_{01}r/a)，
\end{equation}
with $\partial{|\tilde{E}^{(0)}_{\hat{z},\mathrm{ss}}(r,t)|}/\partial t=\partial{\mathbb{A}}/\partial t=0,
$ and  $E_0=-Q\int_V \boldsymbol{J}_z\mathbb{J}_0(x_{01}r/a)dV/[\varepsilon_0\omega_0\pi a^2h\mathbb{J}_0^2(x_{01})]$ being the steady-state electric-field amplitude along the cavity central axis, and $\boldsymbol{J}_z$ the $\hat{z}$ component of the external current $\boldsymbol{J}$ for the excition. It describes a dynamical process wherein the excitation energy of the cavity mode equals to its losses, producing the steady-state response of the dark photons with the amplitude $|\tilde{E}^{(0)}_{\hat{z},\mathrm{ss}}(r,t)|$. The corresponding time-averaged output power is (see Appendix B for details):
\begin{equation}
\langle P^{(0)}_{\mathrm{out}}\rangle=\frac{\kappa\omega_0\varepsilon_0\pi a^2h}{2Q}\mathbb{J}_1^2(x_{01})E_0^2.
\end{equation}

ii) In-situ enhancement of the detectability of dark photons' response.  In the cavity, dark photons dynamically mix now with the pre-existing steady state $\mathrm{TM}_{010}$ mode photons, rather than the previous vacuum state of the cavity mode. As a consequence, the total electric field strength of the cavity mode can be expressed as ~\cite{2018HZ,2022HZ}:
\begin{equation}
\boldsymbol{E}_{\mathrm{ss}}(\boldsymbol{x},t)=\tilde{E}_{\hat{z},\mathrm{ss}}^{(1)}(r,t)+\tilde{E}^{(0)}_{\hat{z},\mathrm{ss}}(r,t).
\end{equation}
The instantaneous output power of the excited $\mathrm{TM}_{010}$ mode reads
\begin{equation}
\begin{aligned}
P_{\mathrm{out}}&=\kappa\frac{\omega_0}{Q}\int_V\left(\frac{1}{2}\varepsilon_{0}|\boldsymbol{E}_{\mathrm{ss}}(\boldsymbol{x},t)|^{2}\right)dV\\&=P_{\mathrm{out}}^{(0)}+P_{\mathrm{out}}^{(1)}+P_{\mathrm{out}}^{(2)},
\end{aligned}
\end{equation}
where $P_{\mathrm{out}}^{(0)}\propto \mathbb{A}^2\cos^2(\omega_0 t)$ refers to the zero-order response whose time average recovers the zero-order output power shown in Eq.~(8), and $P_{\mathrm{out}}^{(2)}\propto \chi^2\mathbb{B}^2\cos^2(\omega_0 t+\phi_d)$ is the second-order energy response of dark photons, which is proportional to $\chi^2$. While,  
\begin{equation}
\begin{aligned}
P^{(1)}_{\mathrm{out}}&\approx\kappa\frac{\omega_0}{Q}\int_V\left(\varepsilon_{0}|\tilde{E}^{(0)}_{z,\mathrm{ss}}(r,t)\tilde{E}_{\hat{z},\mathrm{ss}}^{(1)}(r,t)|\right)dV\\
&=P_1\left[\cos(\phi_d)+\cos(2\omega_0t+\phi_d)\right],
\end{aligned}
\end{equation}
is the first-order instantaneous response power (which is proportional to the first power of $\chi$) with $P_1=\chi 2\pi \kappa\omega_0\varepsilon_0 E_0 c\sqrt{2\rho_d\mu_0}a^2h \cos\theta\mathbb{J}_1(x_{01})/x_{01}$ being its amplitude.

Evidently, if no pre-excited steady state field exists in the cavity (i.e., $\tilde{E}^{(0)}_{\hat{z},\mathrm{ss}}(r,t) = 0$), both the zero-order and first-order dark photon energy response signals vanish. Only the second-order response described by Eq.~(6) remains, reducing this upgraded detection scheme to the conventional dark photon–vacuum photon mixing detection method (with $\boldsymbol{J}=0$). In that case, the cavity only contains vacuum electromagnetic fluctuations described by Eq.~(4), and the dark photon-induced effective current $\boldsymbol{J}_{\mathrm{eff}}$ merely excites the vacuum field of the mode, producing only the second-order energy response signal shown in Eq.~(6). Therefore, the central contribution of this upgraded scheme is the generation of a first-order energy-response signal that is substantially stronger than the original second-order signal, with its power given by Eq.~(11). Clearly, if this first-order signal is detectable, the achievable detection sensitivity of the upgraded scheme will be significantly higher than that of conventional methods relying solely on second-order energy-response signals.

Notably, it is seen from Eq.~(11) that, the instantaneous power of the first-order energy response signal still contains the stochastic phase $\phi_d$. In general, the statistical average of a stochastic signal is zero. Hence, such a stochastic power signal cannot be measured through direct power detection. Fortunately, a stochastic power signal of the form in Eq.~(11) can be detected by using the usual IQ demodulation technique commonly employed in modern communication systems~\cite{2024QY,2024YL}. The key idea of this technique is to decompose the first-order instantaneous power signal $P_{\text{out}}^{(1)}$ into two equal-amplitude co-phased signals:
$S_1(t)=S_2(t)=P_1\left[\cos(\phi_d) + \cos(2\omega_0 t + \phi_d)\right]/\sqrt{2}$. These two signals are then mixed with local oscillator signals $L_1 = L_0\cos(2\omega_0 t)\sqrt{2}$ and $L_2 =L_0\sin(2\omega_0 t)/\sqrt{2}$, respectively. After applying low-pass filters to remove high-frequency components generated by the mixing process, one obtains two orthogonal baseband signals:
$I(\phi_d) =P_1 L_0 \cos(\phi_d)/4$ and $ Q(\phi_d) =P_1 L_0 \sin(\phi_d)/4$. By then performing vector synthesis, the amplitude of the first-order energy response signal, independent of the stochastic phase $\phi_d$, can be recovered as: $P_1= 4\sqrt{I^2(\phi_d) + Q^2(\phi_d)}/L_0$. In other words, although the phase of the dark photon-induced first-order energy response signal is stochastic, its amplitude remains measurable. This capability allows the upgraded detection scheme to exploit the first-order energy response signal to significantly enhance the achievable detection sensitivity of dark photons.

{\it Significantly enhanced achievable detection sensitivity.---}
Obviously, the strongest noise in the upgraded detection system originates from the zeroth-order steady state excitation signal. However, since it carries no dark photon information, it can be treated as a strong background noise satisfying the conventional Maxwell equations. Therefore, it exhibits the physical characteristics that are distinctly different from those of the first-order electromagnetic response, which carries dark photon information. For example, its wave impedance is given by $Z_0(\omega_0)=\mu_0\tilde{E}^{(0)}_{\hat{z},\mathrm{ss}}(r,t)/\tilde{B}^{(0)}_{\phi}(r,t)=\sqrt{\mu_0/\varepsilon_0}\simeq377\Omega$, where $\tilde{B}^{(0)}{\phi}(r,t)$ denotes the magnetic induction of the zeroth order electromagnetic field. In contrast, for typical parameters $\chi=10^{-18}$, $Q=10^{4}$, and $\omega_0/2\pi=8$ GHz, the effective wave impedance of the first-order response signal carrying dark photon information is only $Z_1(\omega_0)=\mu_0\tilde{E}_{\hat{z},\mathrm{ss}}^{(1)}(r,t)/\tilde{B}^{(0)}_{\phi}(r,t)\simeq 5.42\times 10^{-10}\Omega\ll Z_0(\omega_0)$ \cite{2022HZ}. Thus, through either the usual wave impedance matching filtering techniques or the differential detection method~\cite{2025RC}, the electromagnetic signal $\langle{P^{(0)}_{\mathrm{out}}}\rangle$ that contains no dark photon information can be effectively filtered out, enabling the extraction of the first-order energy response signal, under strong coherent noise, is feasible. In this case, the noise that limits the detectability of the first-order energy signal mainly arises from the environmental thermal noise and the shot noise commonly encountered in weak signal detection.
Compared with the conventional experimental schemes for detecting dark photon–photon kinetic mixing effects in vacuum microwave cavities, the power of the present detected signal is linearly proportional to $\chi$, which is significantly stronger than the previous second-order energy mixing signal with the amplitude being proportional to $\chi^2$. Hence, even at the same weak signal detection level, the achievable detection sensitivity of dark photons can be significantly enhanced accordingly.

Fomally, the signal-to-noise ratio(SNR) for the present first-order power signal can be expressed as~\cite{2024APEX}
\begin{equation}
\mathrm{SNR}=\frac{P_{\mathrm{sign}}}{P_{\mathrm{noise}}}\sqrt{t_{\mathrm{int}}\Delta f},
\end{equation}
where $t_{\mathrm{int}}$ is the integration time, $\Delta f$ the detection bandwidth, and $P_{\mathrm{sign}}$ and $P_{\mathrm{noise}}$ denote the signal-and noise powers, respectively. 
The equivalent noise power reads~\cite{2022RC}
$P_{\mathrm{therm}}=\hbar\omega_0 \Delta f/[e^{\hbar\omega_0/(k_bT)}-1]$,
where $k_b$ is the Boltzmann constant and $T$ the equivalent noise temperature. The shot noise of weak signal originates from its statistical fluctuation noise power is~\cite{2022HZ}:
$P_{\mathrm{shot}}=\sigma_N\cdot\hbar\omega/\tau
=\sqrt{P_{1}\hbar\omega_0/\tau}$,
with $\sigma_N=\sqrt{\bar{N}}$ denoting the fluctuation of the mean photon number, $\bar{N}=P_{1}\tau/(\hbar\omega_0)$, and $\tau$ the detector response time. Furthermore, the fluctuation of the detected first-order power signal can be expressed as
\begin{equation}
P_{\mathrm{F}}=\sqrt{\Big(\frac{\partial P_{\mathrm{out}}^{(1)}}{\partial \tilde{E}^{(0)}_{\hat{z},\mathrm{ss}}}\delta \tilde{E}^{(0)}_{\hat{z},\mathrm{ss}}\Big)^2+\Big(\frac{\partial P_{\mathrm{out}}^{(1)}}{\partial \tilde{E}^{(1)}_{\hat{z},\mathrm{ss}}}\delta \tilde{E}^{(1)}_{\hat{z},\mathrm{ss}}\Big)^2},\end{equation}
wherein $\delta \tilde{E}^{(0)}_{\hat{z},\mathrm{ss}}$ and $\delta \tilde{E}^{(1)}_{\hat{z},\mathrm{ss}}$ originated from the fluctuations of the zeroth-order- and first-order electric fields, respectively.
Typically, with the detection duration $t_{\mathrm{int}}=22.1$ s, and detection bandwidth $\Delta f=20$ Hz, the detection confidence level for the present dark photons detection can exceed $95\%$ for the SNR threshold of $\mathrm{SNR}=2$~\cite{2024APEX}.
\begin{figure}[htbp]
\centering
\includegraphics[width=6cm]{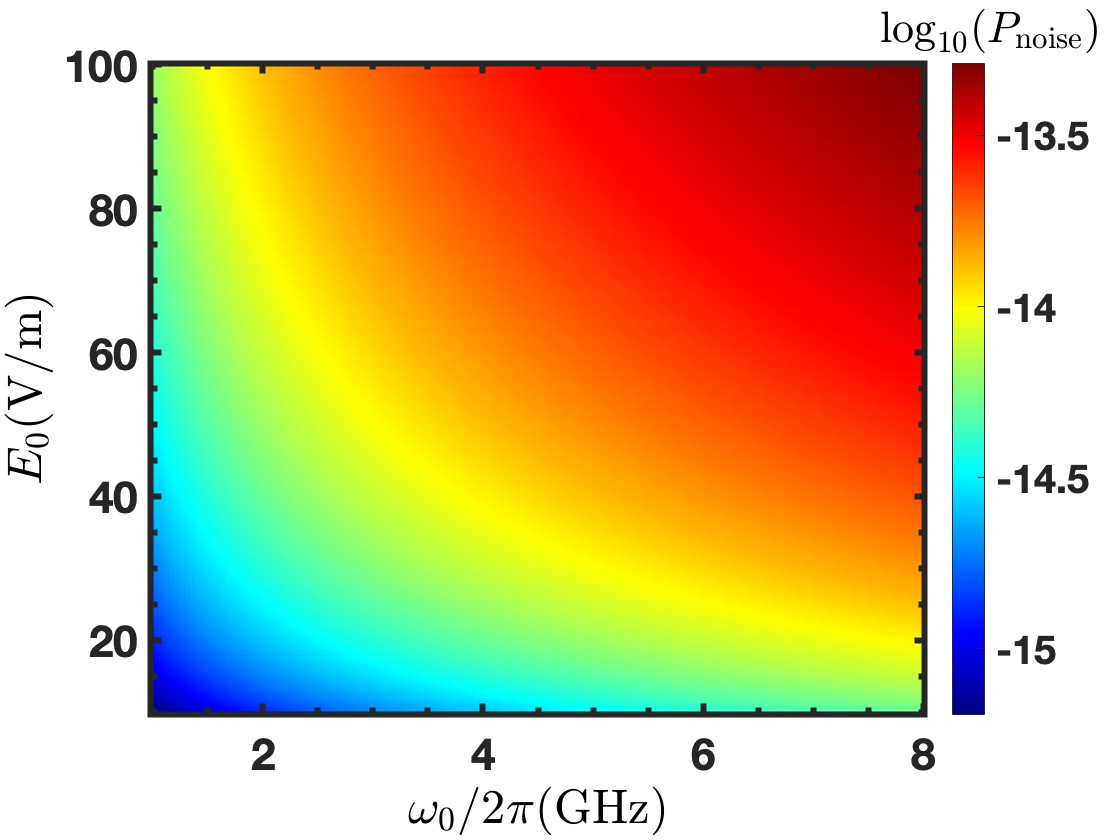}
\caption{The total noise power of the first-order energy response signal versus the resonant frequency $\omega_0/2\pi$ and the pre-excited steady-state field amplitude $E_0$. The relevant parameters are set as: $Q=10^4$, $\rho_{d}=0.45\,\mathrm{GeV/cm^3}$, $\kappa=0.5$, $a=x_{01}\cdot c/\omega_0$, $h=0.9a$, $(\cos\theta)^2=1/3$, $\mathbb{J}_1(x_{01})=0.519$, $\Delta f=20\,\mathrm{Hz}$, $\tau=10^{-3}\,\mathrm{s}$, and $T=10\,\mathrm{mK}$.}
\end{figure}
Specifically, Fig.~2 illustrates the dependence of the total noise power on the resonance frequency $\omega_0/2\pi$ and the steady-state seed field amplitude $E_0$ for the present detection system with the background temperature of $10\,\mathrm{mK}$ and the cavity quality factor $Q=10^4$. One can see that, at the given resonance frequency, the noise power increases with $E_0$. For a fixed $E_0$ it grows with increasing $\omega_0/2\pi$. It should be noted that, in order to preserve the validity of the coherence assumption, the detector response time $\tau$ used in evaluating the shot noise must not exceed the dark photon coherence time $t_{d}$, which is given by $t_{d}=2\pi/(m_{d}v_{d}^{2})$, with $v_{d}\sim 10^{-3}$ being the characteristic relative velocity of dark matter~\cite{2024AI}. Taking representative parameter values of $E_0=10\mathrm{V/m}$, $Q=10^4$, $\tau=10^{-3}\mathrm{s}$, $T=10\mathrm{mK}$, $\Delta f=20\mathrm{Hz}$, and $\omega_0/(2\pi)=8\mathrm{GHz}$, the computed noise powers are $P_{\mathrm{therm}}\approx 2.76\times 10^{-24}\mathrm{W}$ for thermal noise, $P_{\mathrm{shot}}\approx 2.49\times 10^{-18}\mathrm{W}$ for shot noise, and $P_{\mathrm{F}}\approx 1.82\times 10^{-15}\mathrm{W}$ for the fluctuation noise of the first-order signal. Even when the system is operated at room temperature ($300\mathrm{K}$), the thermal noise power only rises to $P_{\mathrm{therm}}\approx 8.28\times 10^{-20}\mathrm{W}$. This indicates that, within the considered parameter regime, the fluctuation noise of the first-order signal dominates the total noise budget, rendering the noise power $P_{F}$ is obviously insensitive to the thermal noise of the detection system.

Based on the usual detectable condition, i.e., $\mathrm{SNR} \geq 1$, and Eq.~(12), the theoretically achievable detection sensitivity for dark photons with the upgraded configuration can be derived now as
\begin{equation}
\begin{aligned}
\chi=&2.11\times 10^{-19}\mathrm{W}\times\frac{P_{\mathrm{noise}}}{1.82\times 10^{-16}\mathrm{W}}\Big(\frac{22.1\mathrm{s}}{t_{\mathrm{int}}}\Big)^{\frac{1}{2}}\Big(\frac{20\mathrm{Hz}}{\Delta f}\Big)^{\frac{1}{2}}\\
\times&\Big(\frac{\omega_0}{1\mathrm{GHz}}\Big)^2\Big(\frac{1\mathrm{V/m}}{E_0}\Big).
\end{aligned}
\end{equation}
In Fig.~3 we plot how the achievable detection sensitivity depends on the steady-state seed field amplitude $E_0$ along the central axis of the cylindrical cavity for the typical parameters;  $Q=6\times10^5$, $\omega_0/(2\pi)=8\mathrm{GHz}$, $t_{\mathrm{int}}=22.1\mathrm{s}$, and $\Delta f=20\mathrm{Hz}$~\cite{2024APEX}. 
\begin{figure}[htbp]
\centering
\includegraphics[width=6cm]{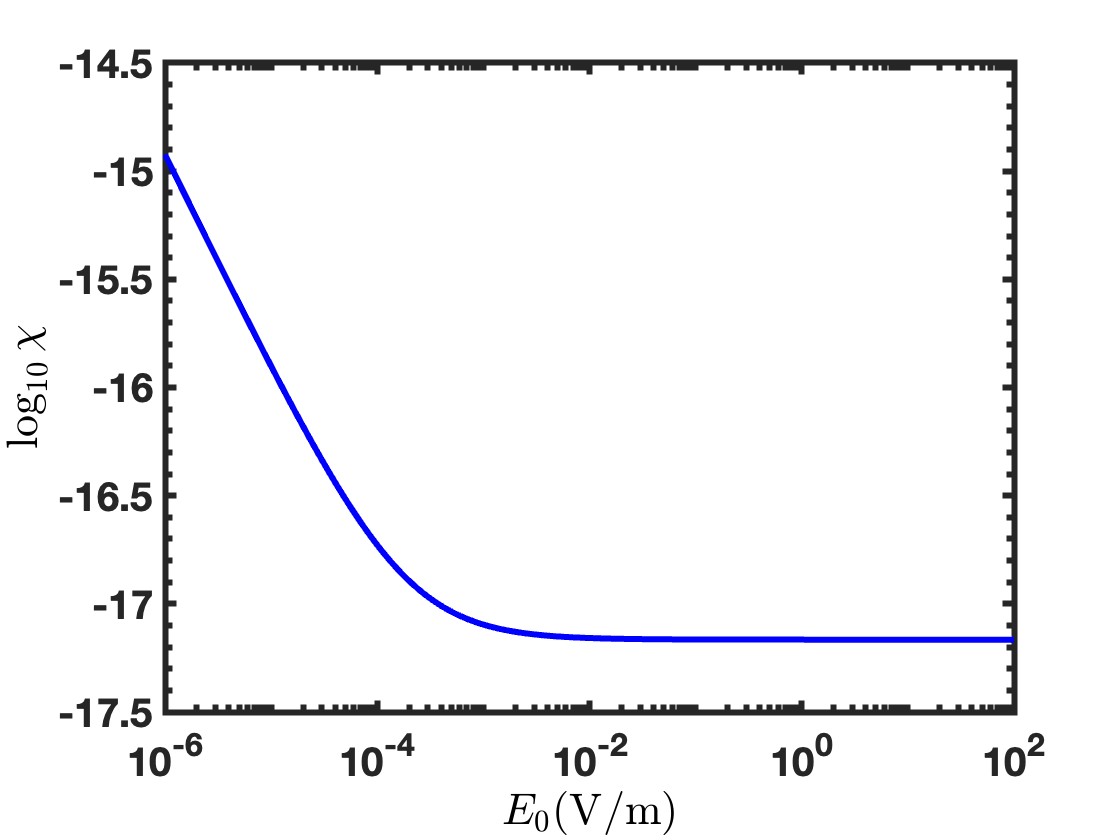}
\caption{The achievable detection sensitivity $\chi$, obtained by detecting the first-order energy response signal, as a function of the pre-excited steady-state field amplitude $E_0$ for $Q=6\times10^5$, $\omega_0/2\pi=8$GHz, $t_{\mathrm{int}}=22.1$s, and $\Delta f_0=20$Hz~\cite{2024APEX}. The remaining parameters are the same as those in Fig.~2.}
\end{figure}
As illustrated, with the increase of $E_0$, the achievable detection sensitivity improves rapidly at first, and then gradually approaches a saturated value belowed at $10^{-17}$ level. Specifically, for $E_0=1\mathrm{V/m}$ the derivative $d\chi/dE_0$ approaches zero, indicating that the sensitivity no longer changes appreciably with further increase of the seed field strength. In this case, the minimal detectable power amplitude of the first-order energy response signal is $P_1^{\mathrm{min}}=2.89\times10^{-19}\mathrm{W}$, corresponding to the mean photon number is $\bar{N}=P_1^{\mathrm{min}} t_{\mathrm{int}}/(\hbar \omega_0 \gamma)\approx 14$ for the detection integration duration of $\sim 22.1$s and the cavity dissipation rate of $\gamma=\omega_0/Q\sim 83776$Hz ~\cite{2024APEX}. Obviously, such a weak signal is well detectable for the current weak microwave detection technique~\cite{2016KI,2018JCB}. Comparatively, to achieve the detection sensitivity of $\chi\approx 1.54\times10^{-16}$ with the usual scheme without any pre-excited field, the power of second-order response signal should be less than $\langle P_{\mathrm{out}}^{(2)}\rangle = 4.97\times10^{-26}\mathrm{W}$, which corresponds to detect the signal with the mean photon number being $\bar{N} \ll 1$. These results clearly demonstrate that, compared to the second-order energy response arising from kinetic mixing between dark photons and vacuum state photons inside the cavity, the present detection scheme, based on a steadily driven microwave cavity, substantially enhances the detectability of dark photons.

\begin{figure}[htbp]
\centering\includegraphics[width=8cm]{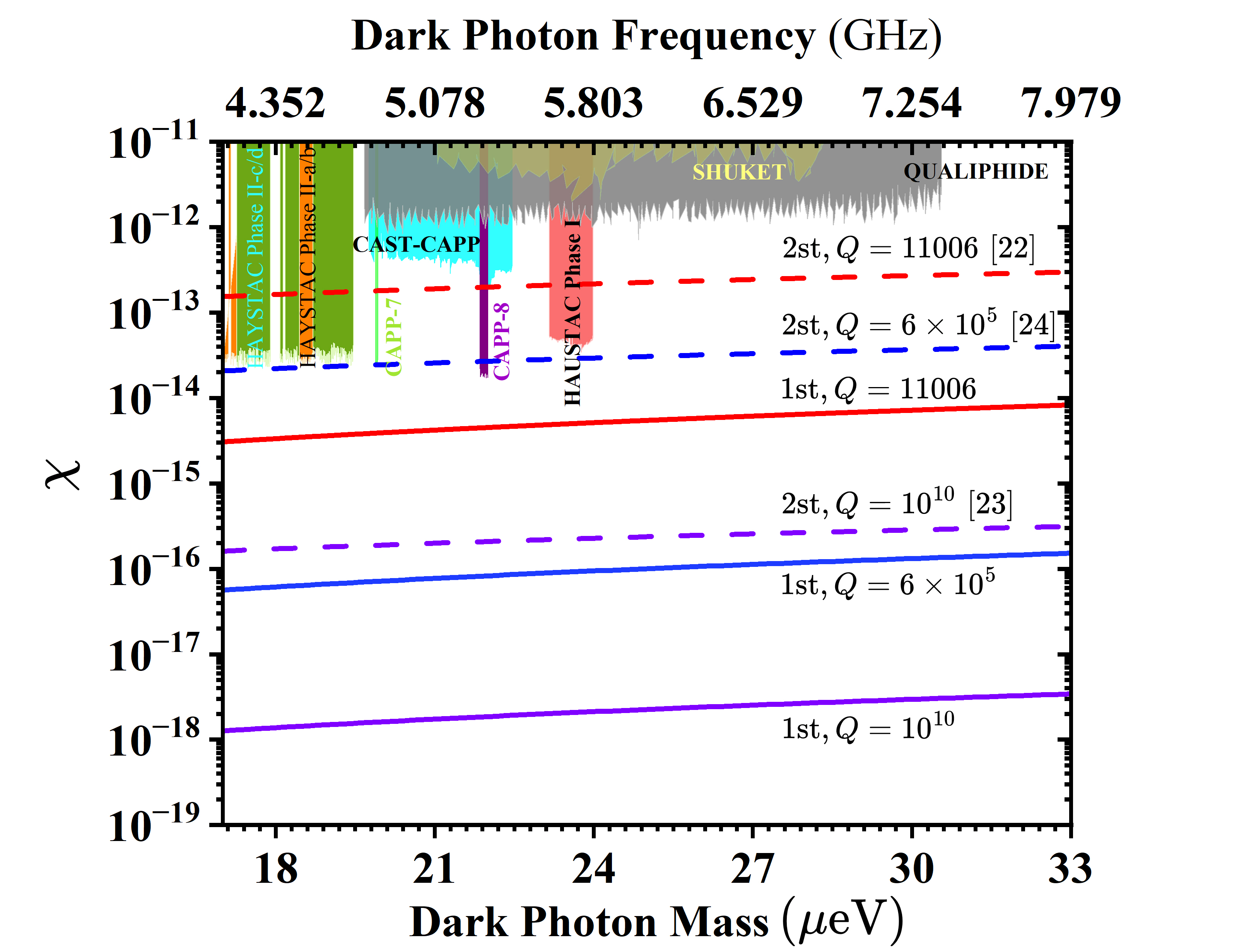}
\caption{
Detection sensitivity $\chi$ as a function of the scan frequency for different quality factors ($Q=11006$, $6\times10^5$, and $10^{10}$). The solid and dashed curves correspond respectively to the sensitivities achieved by the first-order and second-order energy response signals detection, for $E_0=1\mathrm{V/m}$ and $t_{\mathrm{int}}=22.1\mathrm{s}$. The remaining parameters are identical to those in Fig.~2. The colored shaded regions are based on the experimental exclusion limits reported in Ref.~\cite{2020CO}.
}
\end{figure}

To further demonstrate the detectable enhancement with the present scheme, in Fig.~4 we compare specifically its achievable detection sensitivity with those of the usual schemes without any pre-excited field for the cavities with different quality factors $Q$ at $10\,\mathrm{mK}$. One can see clearly that, analogous to the usual second-order energy response signal detections, the achievable detection sensitivity of the present first-order energy response signal detection improves also markedly with the increase of cavity's quality factor. Particularly, for the cavity with the same quality factor $Q$, the achievable detection sensitivity by the first-order energy response signal detection is significantly superior to those achieved by the usual second-order energy response signal detection. Specifically, for $\mathrm{SNR}>2$ (corresponding to a confidence level $>95\%$) and $Q=11006$, the achievable detection sensitivity by the present first-order energy response detection can reach $10^{-15}$ (red solid curve), which is notably better than that of the usual second-order energy response signal detection, which is just approximated as $10^{-13}$ (red dashed curve). For example, compared with the typical vacuum cavity detector, typically such as the detector~\cite{2024APEX} (with the cavity at center frequency $\sim 7.139\,\mathrm{GHz}$,and $Q=11006$, and thus the achieved detection sensitivity is $10^{-13}$), the upgraded configuration proposed here can bring the improvement of detection sensitivity about two orders of magnitude. If the quality factor of the cavity can be enhanced as $Q=6\times 10^5$, the achievable detection sensitivity of the present scheme can be further improved to $10^{-17}$ (blue solid curve), providing roughly three orders enhancement of magnitude. It can surpass even the  experimental level (at $\chi\approx 6\times 10^{-16}$)~\cite{2024RK} by approximately one order of magnitude, for $\omega_0/2\pi=6.52014\,\mathrm{GHz}$, and $Q=6\times 10^5$. Notably, for $Q=10^{10}$, the present achievable detection sensitivity can reach the $10^{-18}$-level (purple solid curve), which is about two orders of magnitude better than that by the usual second-order energy response signal detection (purple dashed curve)~\cite{2024ZXT}. 
Importantly, the power of the first-order energy response signal is still at the level of $10^{-21}\,\mathrm{W}$ with the mean photon number being about $5385$ over the integration time. This signal strength lies well within the detectable range of existing microwave single-photon detection technologies~\cite{2016KI,2018JCB}.

{\it Conclusions and discussions---}
In summary, to address the challenge that the second-order energy response signal (whose power is proportional to $\chi^2$), induced by kinetic mixing between dark photons and vacuum fluctuating electromagnetic fields, in a cavity is exceedingly weak and difficult to be detected, in the present work we proposed an in-situ detectable enhanced scheme based on a pre-steady-state excited field (i.e., the target mode had been pre-excited). As a consequence, the first-order energy response signal (whose power is alternatively proportional to $\chi$), rather than the conventional second-order energy response signal generated from the vacuum, can be generated for the detection. Since such a signal is detectable for the current weak microwave IQ-modulated technique, the achievable detection sensitivity can be improved at least one order of magnitude, compared with the usual detector without any pre-excited field. The numerical estimation indicated that, under cryogenic conditions of $10\,\mathrm{mK}$, with a seed field amplitude of $E_0=1\,\mathrm{V/m}$ on the central axis of a cylindrical cavity with the quality factor of $Q=11006$, the achievable detection sensitivity of the present upgraded scheme enables effective scanning for microwave-band dark photons at a confidence level exceeding $95\%$ and achieve a sensitivity of $10^{-15}$, which is markedly superior to the $\chi\sim 10^{-13}$ level attainable via the usual second-order energy response signal detection. It is worth to emphasize the in-situ detectability enhancement scheme proposed here differs fundamentally from the previous ex-situ amplification, wherein the second-order signal from a vacuum cavity is coherently amplified after emission~\cite{2023ZO}. Such an ex-situ coherent amplification method merely improves the SNR for detecting the second-order effect, whereas the directly measurable quantity, in a statistical average sense, remains the second-order signal. Consequently, the achievable dark photon sensitivity is not intrinsically improved.

Certainly, the primary experimental challenge of the upgraded scheme lies in extracting the relatively weak first-order energy response from the dominant zeroth-order background. Fortunately, this zeroth-order signal is completely deterministic in nature and can be directly subtracted from the acquired data. Therefore, it is not the fundamental limitation to the theoretically attainable sensitivity. Finally, by employing a tuning rod to continuously adjust the cavity mode frequency, the present scheme can also be utilized for the broadband scanning detection of the dark photon--photon mixing effect. For example, with an effective tuning bandwidth of approximately $0.5\,\mathrm{GHz}$~\cite{2025GC} and a multi-detector frequency-division parallel measurement strategy~\cite{2025RQK}, this scheme holds the potential to significantly reduce the total scan time while maintaining high sensitivity, thereby enhancing the efficiency for the broadband searches of dark photons.

\section*{Supplemental material}
The supplemental material includes two parts, providing the relevant derivations in the text.

\begin{center}
	\textbf{A: Derivation of Eq.~(6).}
\end{center}
Let $\boldsymbol{A}_j^\mathrm{cav}(\boldsymbol{x})$ denote the $j$-th source-free eigenmode of the cavity, which satisfies the standard Helmholtz equation
\renewcommand\theequation{A1}
\begin{equation}
-\nabla^2\boldsymbol{A}_j^\mathrm{cav}(\boldsymbol{x})=\frac{\omega_j^2}{c^2}\boldsymbol{A}_j^\mathrm{cav}(\boldsymbol{x}),
\end{equation}
and the orthonormality condition
$\int_V \boldsymbol{A}_j^{\mathrm{cav}}(\boldsymbol{x})\cdot\boldsymbol{A}_{j'}^{\mathrm{cav}}(\boldsymbol{x})\,dV=C_j\delta_{jj'}$ with $C_j$ being a normalization constant. Now, the effective current $\boldsymbol{J}_{\mathrm{eff}}=m_{d}^2\chi\boldsymbol{A}'(\boldsymbol{x},t)/c^2$, arising from kinetic mixing between the dark photon field $\boldsymbol{A}'(\boldsymbol{x},t)$ and the vacuum fluctuating electromagnetic field $\widetilde{\boldsymbol{A}}(\boldsymbol{x},t)$, generates the  electromagnetic response signal $\boldsymbol{A}^{(1)}(\boldsymbol{x},t)$ in the  microwave cavity with eigenfrequency $\omega_j$, 
\renewcommand\theequation{A2}
\begin{equation}
\left(\frac{1}{c^{2}}\frac{\partial^{2}}{\partial t^{2}}-\nabla^{2}\right)\boldsymbol{A}^{(1)}(\boldsymbol{x}, t)=\frac{m_{d}^2}{c^2}\chi \boldsymbol{A}'(\boldsymbol{x},t),
\end{equation}
which is the first-order correction $\boldsymbol{A}^{(1)}(\boldsymbol{x},t)$ to the ordinary electromagnetic field induced by dark photon kinetic mixing. Expanding it in terms of the cavity eigenmodes, i.e.,
$
\boldsymbol{A}^{(1)}(\boldsymbol{x},t)=\sum_j\alpha^{(1)}_j(t)\boldsymbol{A}_j^{\mathrm{cav}}(\boldsymbol{x}),
$
we have
\renewcommand\theequation{A3}
\begin{equation}
\ddot{\alpha}^{(1)}_j(t)+\omega_j^2\alpha^{(1)}_j(t)=\frac{\chi m_{d}^2}{C_j}\int_V \boldsymbol{A}_j^{\mathrm{cav}}(\boldsymbol{x})\cdot\boldsymbol{A}'(\boldsymbol{x},t)\,dV.
\end{equation}
For ultralight wave-like dark photons, the field can be treated as spatially isotropic~\cite{2023JG} and thus $\boldsymbol{A}'(\boldsymbol{x},t)=\boldsymbol{A}'(\boldsymbol{x})e^{-i(m_{d} t+\phi_d)}$ with
\renewcommand\theequation{A4}
\begin{equation}
\boldsymbol{A}'(\boldsymbol{x})=\hat{\boldsymbol{n}}\,\frac{c\sqrt{2\rho_{d}\mu_0}}{m_{d}}.
\end{equation}
Here, $\rho_d$ is the local energy density of the dark photon field near the Earth, $\phi_d$ is a stochastic phase, $\hat{\boldsymbol{n}}$ denotes the polarization direction of the dark photon, and $\mu_0$ is the vacuum permeability~\cite{2024AI}.

In practice, the microwave resonant cavity exhibits dissipation, and thus its quality factor $Q$ should be finite. Therefore, Eq.~(A3) must be modified to include a damping term $\omega_j/Q$, such that the response of the effective current $\boldsymbol{J}_{\mathrm{eff}}$ can be  modified as
\renewcommand\theequation{A5}
\begin{equation}
\left(\frac{d^{2}}{dt^{2}}+\frac{\omega_{j}}{Q}\frac{d}{dt}+\omega_{j}^{2}\right)\alpha^{(1)}_{j}(t)=b^{(1)}_{j}e^{-i(m_{d} t+\phi_d)}
\end{equation}
where $b^{(1)}_j=\chi m_{d}^2\int_V [\boldsymbol{A}_j^{\mathrm{cav}}(\boldsymbol{x})\cdot\boldsymbol{A'}(\boldsymbol{x})/C_j]dV$ represents the driving source of  the mode. Clearly, Eq.~(A5) admits a steady state solution with the form $\alpha^{(1)}_j(t)=\alpha^{(1)}_{j,0}e^{-i(m_{d} t+\phi_d)}$ and 
\renewcommand\theequation{A6}
\begin{equation}
\alpha^{(1)}_{j}(t)=\frac{b^{(1)}_j}{\omega_j^2-m_{d}^2-im_{d}\omega_0/Q}e^{-i(m_{d} t+\phi_d)}.
\end{equation}
This indicates that the effective current $\boldsymbol{J}_{\mathrm{eff}}$ can excite a steady-state field of the $j$th mode inside the cavity with the strength
\renewcommand\theequation{A7}
\begin{equation}
\boldsymbol{\tilde{E}}_j^{(1)}(\boldsymbol{x},t)=-\dot{\boldsymbol{A}}_j^{(1)}(\boldsymbol{x},t)=im_{d}\alpha^{(1)}_{j,0}e^{-i(m_{d} t+\phi_d)}\boldsymbol{A}_{j}^{\mathrm{cav}}(\boldsymbol{x}),
\end{equation}
where $\boldsymbol{A}_j^{(1)}(\boldsymbol{x},t)=\alpha^{(1)}_j(t)\boldsymbol{A}_j^{\mathrm{cav}}(\boldsymbol{x})$. Obviously, the power of such a response signal reads 
\renewcommand\theequation{A8}
\begin{equation}
P^{(2)}_{\mathrm{out}}\approx\kappa\frac{\omega_j}{Q}\int\left(\frac{1}{2}\varepsilon_{0}|\boldsymbol{\tilde{E}}_j^{(1)}(\boldsymbol{x},t)|^{2}\right)dV,
\end{equation}
with $\kappa$ denoting the cavity-detector coupling constant~\cite{2012PA}.
Accordingly, for the resonant response  (i.e., $m_d=\omega_j$), the time-averaged response power can be expressed as~\cite{2012PA}
\renewcommand\theequation{A9}
\begin{equation}
\begin{aligned}
\langle P_{\mathrm{out}}^{(2)}\rangle \propto&\frac{1}{\tau}\int_{0}^{\tau}[\boldsymbol{\tilde{E}}_j^{(1)}(\boldsymbol{x})]^2\operatorname{cos}^{2}(\omega_j t+\phi_d)dt\\
=&\kappa\chi^{2}\omega_j\rho_d QV\mathrm{G},
\end{aligned}
\end{equation}
where $\mathrm{G}=\left|\int_V \boldsymbol{A}_j^{\mathrm{cav}}(\boldsymbol{x})\cdot\hat{\boldsymbol{n}}dV\right|^{2}/(V\int_V \left|\boldsymbol{A}_j^{\mathrm{cav}}(\boldsymbol{x})\right|^{2}dV)$ is the geometry form factor of the mode~\cite{2024APEX,2024RK}.

Specifically, for the $\mathrm{TM}_{010}$ mode in the cavity, Eq.~(A6) reduces to $\alpha_{010,0}^{(1)}=iQb_{010}^{(1)}/\omega_{0}^{2}$ and Eq.~(A7) can be rewritten as
\renewcommand\theequation{A10}
\begin{equation}
\boldsymbol{\tilde{E}}_{\mathrm{ss}}^{(1)}(\boldsymbol{x})=-\frac{\chi Q \omega_0\int_{V}\boldsymbol{A}_{010}^{\mathrm{cav}}(\boldsymbol{x})\cdot\boldsymbol{A}^{\prime}(\boldsymbol{x})dV}{\int_{V}|\boldsymbol{A}_{010}^{\mathrm{cav}}(\boldsymbol{x})|^{2}dV}\boldsymbol{A}_{010}^{\mathrm{cav}}(\boldsymbol{x}),
\end{equation}
for the resonant response, i.e., $m_{d}=\omega_{010}=\omega_0$. As the $\mathrm{TM}_{010}$ mode in a cylindrical cavity possesses only an electric field along the axial $\hat{z}$ direction, its vector potential can be specifically expressed as $\boldsymbol{A}_{010}^{\mathrm{cav}}(r)=\hat{z}A_0\mathbb{J}_0(x_{01}r/a)$, where $A_0$ is the amplitude, $\mathbb{J}_0(\cdot)$ is the zero-order Bessel function of the first kind, $x_{01}\approx 2.4048$ is the first root of $\mathbb{J}_0(\cdot)$, $r$ denotes the radial coordinate, and $a$ is the radius of the cylindrical cavity. Hence, for the $\mathrm{TM}_{010}$ mode we have 
$\int_V|\boldsymbol{A}_{010}^{\mathrm{cav}}(r)|^2dV=|A_0|^2\pi ha^2\mathbb{J}_1^2(x_{01})$ and $\int_V\boldsymbol{A}_{010}^{\mathrm{cav}}(r)\cdot\boldsymbol{A}'(\boldsymbol{x})dV=A_0 2\pi hc\sqrt{2\rho_{d}\mu_0}a^2\mathbb{J}_1(x_{01})\cos\theta/(\omega_0x_{01})$. Here, $\cos\theta=\langle\hat{z},\hat{\boldsymbol{n}}\rangle$ is the angle between the dark photon polarization direction and the cavity $\hat{z}$-axis, $\mathbb{J}_1(\cdot)$ is the first-order Bessel function of the first kind, and $h$ is the height of the cylindrical cavity. Consequently, driven resonantly by the effective current, a steady-state field in the $\mathrm{TM}_{010}$ mode can be  excited as
\renewcommand\theequation{A11}
\begin{equation}
\tilde{E}_{\hat{z},\mathrm{ss}}^{(1)}(r,t)
=-\chi\frac{2 Qc\sqrt{2\rho_d\mu_0}\mathbb{J}_0\left(\frac{x_{01}}{a}r\right)\cos\theta}{x_{01}\mathbb{J}_1(x_{01})}e^{-i(\omega_0 t+\phi_d)}.
\end{equation}
Its time-averaged power, i.e., Eq.~(A9), can be written explicitly as
\renewcommand\theequation{A12}
\begin{equation}
\begin{aligned}
\langle P_{\mathrm{out}}^{(2)}\rangle\approx&\kappa\frac{\omega_0}{Q}\int_V\left(\frac{1}{2}\varepsilon_{0}\Big|\tilde{E}_{\hat{z},\mathrm{ss}}^{(1)}(r,t)\Big|^{2}\right)dV\\
=&\chi^2\frac{4\kappa\omega_0 Q\rho_d\pi h a^2\cos^2\theta}{x^2_{01}}.
\end{aligned}
\end{equation}
This is precisely Eq.~(6) in the main text.

\begin{center}
	\textbf{B: Derivation of Eq.~(7).}
\end{center}

Following Ref.~\cite{2011DMP}, the steady-state electric field of the $\mathrm{TM}_{010}$ mode in the cavity, under the driving of a source $\boldsymbol{J}(\boldsymbol{x},t)$,  can be simply expressed as
\renewcommand\theequation{B1}
\begin{equation}
\tilde{E}^{(0)}_{\hat{z},\mathrm{ss}}(r,t)=E_0 \mathbb{J}_0\left(\frac{x_{01}}{a}r\right)e^{-i\omega_0 t},
\end{equation}
where $E_0=-Q\int_V \boldsymbol{J}_z\mathbb{J}_0(\frac{x_{01}r}{a})dV/[\varepsilon_0\omega_0\pi a^2h\mathbb{J}_0^2(x_{01})]$ is the amplitude of the $\mathrm{TM}_{010}$ electric field along the central axis of the cavity, and $\boldsymbol{J}_z$ denotes the $z$-component of the ordinary electromagnetic current. The time-averaged energy density of such a steady-state excited mode is
\renewcommand\theequation{B2}
\begin{equation}
\begin{aligned}
\langle w^{(0)}_e(r,t)\rangle=&\frac{1}{\tau}\int_0^\tau\frac{1}{2}\varepsilon_0E_0^2\mathbb{J}^2_0\left(\frac{x_{01}}{a}r\right)\cos^2\omega_0t dt\\
=&\frac{\varepsilon_0}{4}E_0^2\mathbb{J}^2_0\left(\frac{x_{01}}{a}r\right),
\end{aligned}
\end{equation}
and thus the total energy stored in the excited mode reads
\renewcommand\theequation{B3}
\begin{equation}
\begin{aligned}
U^{(0)}&=2\int_V\langle w^{(0)}_e(r,t)\rangle dV\\
&=\frac{\varepsilon_0}{2}\int_0^hdz\int_0^{2\pi}d\varphi\int_0^a r E_0^2\mathbb{J}_0^2\left(\frac{x_{01}}{a}r\right)dr\\
&=\frac{\varepsilon_0}{2}E_0^2\pi a^2h \mathbb{J}_1^2(x_{01}).
\end{aligned}
\end{equation}
Hence, the time averaged power of such a zeroth-order response signal of dark photons is 
\renewcommand\theequation{B4}
\begin{equation}
\langle P^{(0)}_{\mathrm{out}}\rangle=\kappa\frac{\omega_0 U^{(0)}}{Q}=\frac{\kappa\omega_0\varepsilon_0\pi a^2h}{2Q}\mathbb{J}_1^2(x_{01})E_0^2.
\end{equation}
This is just Eq.~(7) in the main text.

{\it Acknowledgments---}
This work was partially supported in part by the National Key Research and Development Program of China under Grant No.~2021YFA0718803, the National Natural Science Foundation of China under Grant No.~P110325G02011, No.~12505083, the Fundamental Research Funds for the Central Universities under Grant No.~2682024CX048, and the Natural Science Foundation of Sichuan under Grant No.~2025ZNSFSC0857.

{\it Date availability---}
The data are not publicly available. The data are available from the authors upon reasonable request.


\begin{thebibliography}{20}
\bibitem{2018RA}R. Aaij, B. Adeva, M. Adinolfi, et al. (LHCb Collaboration), Search for Dark Photons Produced in 13 TeV $\textit{pp}$ Collisions, Phys. Rev. Lett. {\bf 120}, 061801 (2018).
\bibitem{2025TL}T. Li, Z. H. Bo, W. Chen, et al., Search for MeV-Scale axionlike particles and dark photons with PandaX-4T, Phys. Rev. Lett. {\bf 134}, 071004 (2025).
\bibitem{2017IMB}I. M. Bloch, R. Essig, K. Tobioka, et al., Searching for dark absorption with direct detection experiments, J. High Energ. Phys. {\bf 2017}, 87 (2017).
\bibitem{2025YF}Y. Fang, C. Gao, Y. Y. Li, J. Shu, Y. S. Wu, H. X. Xing, B. Xu, L. L. Xu, and Chen Zhou, Quantum frontiers in high energy physics. Sci. China Phys. Mech. Astron. {\bf 68}, 260301 (2025). 
\bibitem{2022DA}D. Antypas, A. Banerjee, C. Bartram, et al., Snowmass 2021 white paper New horizons: scalar and vector ultralight dark matter, arXiv:2203.14915.
\bibitem{2025HF}H. Fukuda, Y. Matsuzaki, and T. Sichanugrist, Directional searching for light dark matter with quantum sensors, Phys. Rev. Lett. {\bf 135}, 241802 (2025).
\bibitem{2019VVF}V. V. Flambaum, I. B. Samsonov1, and H. B. Tran Tan, Interference-assisted detection of dark photon using atomic transitions, Phys. Rev. D {\bf 99}, 115019 (2019).
\bibitem{2020AB}A. Bhoonah, J. Bramante, and N. Q. Song, Superradiant searches for dark photons in two stage atomic transitions, Phys. Rev. D 101, 055040 (2020).
\bibitem{2021AVD}A. V. Dixit et al., Searching for dark matter with a superconducting qubit, Phys. Rev. Lett. {\bf 126}, 141302 (2021).
\bibitem{2023SC}S. Chen, et al., Detecting hidden photon dark matter using the direct excitation of transmon qubits, Phys. Rev. Lett. {\bf 131}, 211001 (2023).
\bibitem{2023JG}J. Gu\'{e}, A. Hees
, J. Lodewyck, R. L. Targat, and P. Wolf, Search for vector dark matter in microwave cavities with Rydberg atoms, Phys. Rev. D {\bf 108}, 035042 (2023).
\bibitem{2024AA}A. Agrawal, A. V. Dixit, T. Roy, et al., Stimulated emission of signal photons from dark matter
waves, Phys. Rev. Lett. {\bf 132}, 140801 (2024).
\bibitem{2025SRH}S. R. He, D. He, Y. F. Li, L. Gao, X. N. Feng, H. Zheng, and L. F. Wei, Sensitively searching for microwave dark photons with atomic ensembles, Phys. Rev. D {\bf 112}, 063536 (2025).
\bibitem{2025SC}S. Chigusa, M. Hazumi, E. D. Herbschleb, N. Mizuochi, and K. Nakayama, Light dark matter search with nitrogen-vacancy centers in diamonds. J. High Energ. Phys. {\bf 2025}, 83 (2025).
\bibitem{2025HS}H. Shi, A. J.Brady, W. G\'{o}recki,  L. Maccone, R. D. Candia, and Q. T. Zhuang, Quantum-enhanced dark matter detection with in-cavity control: mitigating the Rayleigh curse. npj Quantum Inf. {\bf 11}, 48 (2025).
\bibitem{2024AI}A. Ito, R. Kitano, W. Nakano, et al., Quantum entanglement of ions for light dark matter detection, J. High Energ. Phys. {\bf 2024}, 124 (2024).
\bibitem{2024SC}S. Chen, H. Fukuda, T. Inada, T. Moroi, T. Nitta1, and T. Sichanugrist, Quantum Enhancement in Dark Matter Detection with Quantum Computation, Phys. Rev. Lett. {\bf 133}, 021801 (2024).
\bibitem{2012PA}P. Arias, D. Cadamuro, M. Goodsell, J. Jaeckel, J. Redondo, and A. Ringwald, WISPy cold dark matter, 
J. Cosmol. Astropart. Phys. {\bf 2012}, 013 (2012).
\bibitem{2025YFC}Y. F. Chen, C. L. Li, Y. X. Liu, J. Shu, Y. T. Yang, and Y. J. Zeng, Simultaneous resonant and broadband detection of ultralight dark matter and high-frequency gravitational waves via cavities and circuits, Rep. Prog. Phys. {\bf 88}, 057601 (2025).
\bibitem{2021AC}A. Caputo, A. J. Millar, C. A. J. O’Hare, and E. Vitagliano,
Dark photon limits: A handbook, Phys. Rev. D {\bf 104}, 095029 (2021).
\bibitem{2022RC}R. Cervantes, G. Carosi, S. Kimes, et al., ADMX-Orpheus first search for 70$\mu$⁢eV dark photon dark matter: Detailed design, operations, and analysis, Phys. Rev. D {\bf 106}, 102002 (2022).
\bibitem{2024APEX}D. He, J. Fan, X. Gao, et al., Dark photon constraints from a 7.139 GHz cavity haloscope experiment, Phys. Rev. D {\bf 110}, L021101 (2024).
\bibitem{2024ZXT}Z. X. Tang, B. Wang, Y. F. Chen, et al., First scan search for dark photon dark matter with a tunable superconducting radio-frequency cavity, Phys. Rev. Lett. {\bf 133}, 021005 (2024). 
\bibitem{2024RK}R. Kang, M. Jiao, Y. Tong, Y. Liu, Y. Zhong, Y.-F. Cai, J. Zhou, X. Rong, J. Du, Near-quantum-limited haloscope search for dark-photon dark matter enhanced by a high-$Q$ superconducting cavity, Phys. Rev. D {\bf 109}, 095037 (2024).
\bibitem{2025YY}Y. Yin, R. Q. Kang, M. Jiao, and X. Rong, Haloscope searching for dark photons in the $Q$ band with a novel coupling tuning structure, Phys. Rev. D {\bf 111}, 095022 (2025).
\bibitem{2024RC}R. Cervantes, J. Aumentado, C. Braggio, et al., Deepest sensitivity to wavelike dark photon dark matter with superconducting radio frequency cavities, Phys. Rev. D {\bf 110}, 043022 (2024).
\bibitem{2025YJZ}Y. J. Zeng, Y. X. Liu, C. L. Li, et al., Cavity as radio telescope for galactic dark photon, Sci. Bull. {\bf 70}, 661 (2025).
\bibitem{2024QY}Q. Yang, Y. Gao, and Z. Peng, Quantum dual-path interferometry scheme for axion dark matter searches. Commun Phys {\bf 7}, 277 (2024).
\bibitem{2024YL}Y. F. Li, S. R. He, M. Zhang, and L. F. Wei, Quantum computation with electrons trapped on liquid Helium by using the centimeter-wave manipulating techniques, Quantum Inf. Process. {\bf 23}, 294 (2024).
\bibitem{2018HZ}H. Zheng, L. F. Wei, H. Wen, and F. Y. Li, Electromagnetic response of gravitational waves passing through an alternating magnetic field: A scheme to probe high-frequency gravitational waves, Phys. Rev. D {\bf 98}, 064028 (2018).
\bibitem{2022HZ}H. Zheng and L. F. Wei, Experimental system to detect the electromagnetic response of high-frequency gravitational waves, Phys. Rev. D {\bf 106}, 104003 (2022).
\bibitem{2025RC}R. Corgier, M. Malitesta, L. A. Sidorenkov, et al., Optimized squeezing for accurate differential sensing under large phase noise, Quantum Sci. Technol. {\bf 10}, 045016 (2025).
\bibitem{2016KI}K. Inomata, Z. Lin, K. Koshino, W. D. Oliver, J. S. Tsai, T. Yamamoto, and Y. Nakamura, Single microwave-photon detector using an artificial $\Lambda$-type three-level system, Nat. Commun. {\bf 7}, 12303 (2016).
\bibitem{2018JCB}J. C. Besse, S. Gasparinetti, M. C. Collodo, T. Walter, P. Kurpiers, M. Pechal, C. Eichler, and A. Wallraff, Single-shot quantum nondemolition detection of individual itinerant microwave photons, Phys. Rev. X {\bf 8}, 021003 (2018).
\bibitem{2020CO}C. O’Hare, cajohare/Dark photon limits, https://cajohare.github.io/AxionLimits/docs/dp.html.
\bibitem{2023ZO}Z. Omarov, J. Jeong, and Y. K. Semertzidis, Speeding axion haloscope experiments using heterodyne-variance-based detection with a power meter, Phys. Rev. D {\bf 107}, 103005 (2023).
\bibitem{2025GC}G. Carosi, C. Cisneros, N. Du, et al., Search for Axion dark matter from 1.1 to 1.3 GHz with ADMX, Phys. Rev. Lett. {\bf 135}, 191001 (2025).
\bibitem{2025RQK}
R. Q. Kang, Q. Q. Hu, X. Cai, W. L. Yu, J. W. Zhou, X. Rong, and J. F. Du,
Scalable architecture for dark photon searches: superconducting-qubit proof of principle, Phys. Rev. Lett. {\bf 135}, 181004 (2025).
\bibitem{2011DMP}D. M. Pozar, Microwave Engineering, \textit{Microwave Engineering} (John Wiley and Sons, 2011) pp.~288-297.
\end{thebibliography}
\end{document}